\documentclass{aa}
\usepackage[varg]{txfonts}
\usepackage{graphicx}
\usepackage{epstopdf}
\usepackage{bm}
\usepackage{threeparttable}
\usepackage{multirow}
\usepackage{xcolor}
\begin{document}

\title{Molecule-specific diffusion and desorption of interstellar ices on carbonaceous dust}
\titlerunning{Diffusion and desorption of interstellar ices on carbonaceous dust}
\authorrunning{Chiu et al.}

   \author{Yi-Hsuan Chiu\inst{1}\email{ginachiu@phy.ncu.edu.tw}
   \and Tushar Suhasaria\inst{2}\corrauth{suhasaria@mpia.de}
   \and Cornelia J{\"a}ger\inst{2,3}\email{cornelia.jaeger@uni-jena.de}
   \and Chun-Yi Lee\inst{1}\email{kailee@phy.ncu.edu.tw}
   \and Ko-Ju Chuang\inst{4}\email{chuang@strw.leidenuniv.nl}
   \and Thomas Henning\inst{2}\email{henning@mpia.de}
   \and Yu-Jung Chen\inst{1}\corrauth{asperchen@phy.ncu.edu.tw}} 
   
   \institute{Department of Physics, National Central University, Zhongli Dist., Taoyuan City 320317, Taiwan\\ \and Max Planck Institute f\"ur Astronomie, K\"onigstuhl 17, 69117, Heidelberg, Germany\\ \and Laboratory Astrophysics Group of the Max Planck Institute for Astronomy at the Friedrich Schiller University Jena, Institute of Solid State Physics, Helmholtzweg 3, 07743, Jena, Germany\\ \and Laboratory for Astrophysics, Leiden Observatory, Leiden University, P.O. Box 9513, 2300 RA Leiden, The Netherlands\\ }

\abstract
{Interstellar ices form on dust grains in the coldest regions of molecular clouds and preserve key volatile reservoirs that could be incorporated into protoplanetary disks during star and planet formation. However, the effect of the dust surface composition on the ice structure and spectroscopic behavior remains poorly constrained. We present a comparative laboratory study of astrophysically relevant ices (CO, CO$_2$, and H$_2$O) deposited on inert calcium fluoride (CaF$_2$) substrate and carbonaceous dust analogs under interstellar conditions. Infrared spectroscopy and temperature-programmed desorption reveal pronounced molecule-specific infrared spectral responses to the amorphous carbonaceous surface. CO and CO$_2$ both exhibit broadened absorption bands, redshifted band positions, and delayed desorption, arising from thermally activated diffusion into the porous dust matrix and indicating strong molecule–surface interactions. By contrast, H$_2$O varies only very little spectrally and thermally, indicating weak wetting and limited coupling to the substrate. These results provide direct laboratory evidence that dust-ice interfaces can affect the ice structure and desorption kinetics of interstellar ices even in thick ice layers. These findings offer new constraints for interpreting infrared absorption bands in astronomical observations and highlight the importance of surface effects in models of interstellar ice chemistry.}

\keywords{methods: laboratory: solid state -- techniques: spectroscopic -- ISM: lines and bands -- ISM: molecules -- infrared: ISM}
\maketitle
\nolinenumbers
\section{Introduction}

In cold and dense regions of the interstellar medium, gas-phase atoms and molecules accrete onto submicron-sized siliceous and/or carbonaceous dust grains, building icy mantles that include H$_2$O, CO, and CO$_2$ and promoting the formation of complex organic molecules in molecular clouds and protoplanetary disks \citep{oberg2016}. Their composition and structural evolution are primarily constrained through infrared spectroscopy, where band positions, profiles, and intensities encode information about the ice morphology and the local environment \citep{yang2022, mcclure2023}.

Despite the formation of interstellar ices on grain surfaces, substrate effects are commonly assumed to become negligible when the ice layer is sufficiently thick \citep{burke2010}. Laboratory studies therefore often employ smooth inert substrates to simplify the analysis of the ice structure, infrared band profiles, and desorption behavior \citep{gerakines1995, ehrenfreund1997}. However, real interstellar grains possess porous morphologies and large surface areas \citep{jager2008, fulle2017}. Ice growth under such conditions can deviate from simple layer-by-layer deposition, leaving exposed grain surfaces or regions covered by only a few molecular layers \citep{potapov2020}. Interstellar dust grains can strongly affect ice growth, molecular ordering, and desorption dynamics \citep{potapov2021}. For example, CO$_2$ ice deposited on porous amorphous silicates exhibits infrared band profiles distinct from those observed on smooth substrates \citep{suhasaria2025}. Systematic molecule-resolved studies on realistic dust analogs therefore remain limited, leaving open how surface properties shape observable spectral features.

We employed CaF$_2$ as an infrared-transparent reference substrate, and the carbonaceous dust served as the astrophysically relevant porous grain analog. We present a comparative infrared spectroscopic and thermal desorption study of the most abundant interstellar ices, namely CO, CO$_2$, and H$_2$O, deposited on CaF$_2$ and laboratory-synthesized carbonaceous dust under conditions relevant to the interstellar medium. We identify molecule-specific spectral shifts, band broadening, and delayed desorption caused by interactions with the porous dust surface. These measurements provide direct experimental evidence that substrate effects modulate infrared band profiles. This offers new constraints for interpreting data from the James Webb Space Telescope (JWST) and forthcoming missions targeting the ice inventory and thermal structure of protostellar and protoplanetary environments. It is therefore essential to incorporate surface effects into astrochemical models to capture the full diversity of the ice behavior in astrophysical settings.

\section{Experiments}

We conducted experiments using an interstellar photoprocess system. The chamber, equipped with a closed-cycle helium cryostat and a tunable sample heater, reached a base pressure of 5$\times$10$^{-10}$ Torr and allowed us to control the temperature between 15 and 300 K \citep{chen2014}.

Amorphous carbonaceous dust grains were prepared via pulsed laser ablation of graphite in a helium atmosphere and deposited onto CaF$_2$ substrates under high vacuum following the procedure of \citet{jager2008}, who characterized the morphology and porosity of these carbonaceous dust analogs in detail. The deposited carbon layer was approximately 100 nm thick and exhibited a porosity of about 90\%. Before the ice deposition, the carbonaceous dust analogs were vacuum ultraviolet-irradiated following the cleaning procedure of \citet{chuang2023} to remove atmospheric adsorbates formed during air exposure without modifying the chemical or structural properties of the dust.

Gas-phase CO, CO$_2$, and H$_2$O were introduced through stainless-steel lines, with a background pressure maintained below 1$\times$10$^{-7}$ Torr. The ices were deposited at 15 K onto either a bare CaF$_2$ substrate or a pre-coated amorphous carbonaceous dust. Temperature-programmed desorption (TPD) experiments were performed using a constant heating rate of 0.5 K min$^{-1}$. The ice evolution during deposition and warm-up was monitored in situ via a Fourier transform infrared spectrometer at an incident angle of 45\degr. Infrared spectra were taken with a resolution of 1 cm$^{-1}$ for CO and 2 cm$^{-1}$ for CO$_2$ and H$_2$O. The column densities were derived from integrated infrared absorbance using band strengths of 1.1$\times$10$^{-17}$, 7.6$\times$10$^{-17}$, and 1.5$\times$10$^{-16}$ cm molecule$^{-1}$ for the CO $\nu_1$, CO$_2$ $\nu_3$, and H$_2$O stretching regions, respectively \citep{bouilloud2015}.

\section{Results and discussion}

\begin{figure}
\resizebox{\hsize}{12.4cm}{\includegraphics{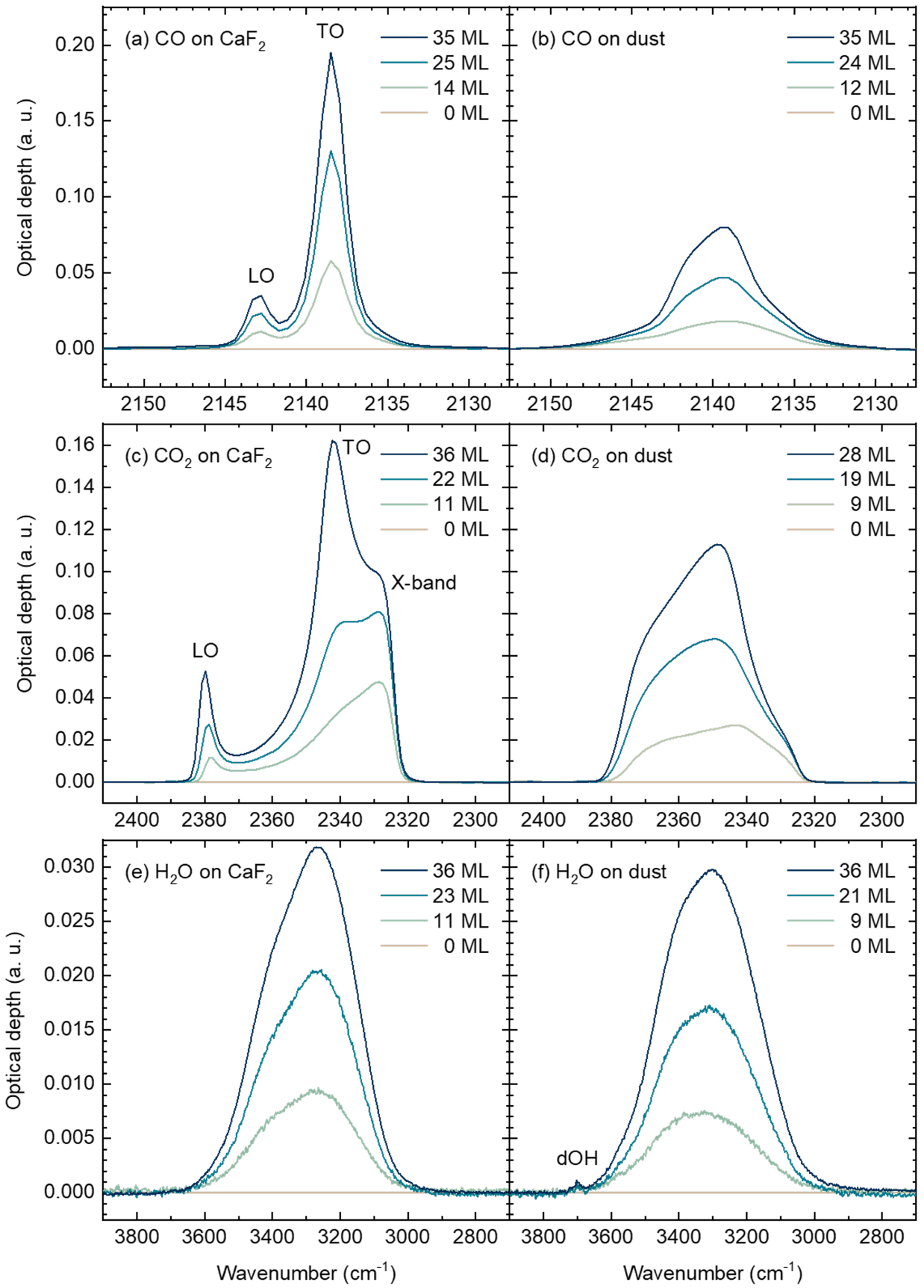}}
\caption{Infrared absorption spectra of CO (top row: a, CaF$_2$; b, carbonaceous dust), CO$_2$ (middle row: c, CaF$_2$; d, carbonaceous dust), and H$_2$O (bottom row: e, CaF$_2$; f, carbonaceous dust) ices at 15 K.}
   \label{fig2}
\end{figure}

\begin{figure}
\resizebox{\hsize}{12.4cm}{\includegraphics{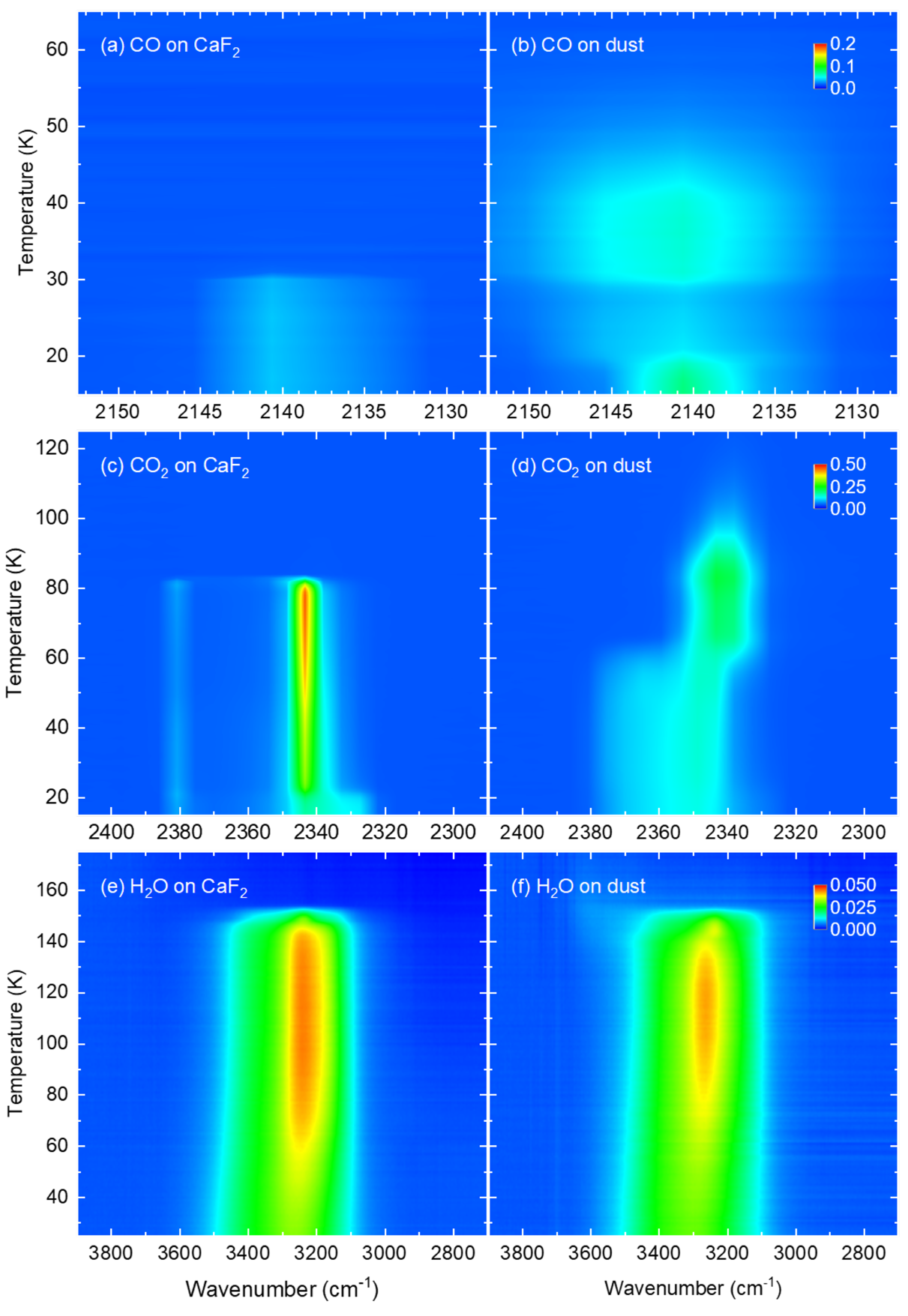}}
\caption{Thermal evolution of CO (top row: a, CaF$_2$; b, carbonaceous dust), CO$_2$ (middle row: c, CaF$_2$; d, carbonaceous dust), and H$_2$O (bottom row: e, CaF$_2$; f, carbonaceous dust) ices. The color scale reflects the relative optical depth of the spectra acquired during thermal processing.}
   \label{fig3}
\end{figure}

\citet{gerakines2023} showed that CO forms spectroscopically well-defined $\alpha$-CO crystalline domains, whereas CO$_2$ and H$_2$O are initially deposited in amorphous forms at 15 K. The following sections compare their substrate-dependent infrared absorption bands and thermal evolution.

\subsection{Carbon monoxide}

In the CO stretching ($\nu_1$) region, the spectra measured on CaF$_2$ (Fig.~\ref{fig2}a) display two sharp peaks at 2143 and 2139 cm$^{-1}$, corresponding to longitudinal-optical (LO) and transverse-optical (TO) vibrational modes \citep{palumbo2006}. The invariance of the peak positions during deposition indicates a stable local environment and minimum molecule–surface coupling.

By contrast, the spectra obtained on carbonaceous dust (Fig.~\ref{fig2}b) exhibit a broadened absorption band centered near 2139 cm$^{-1}$ with a shoulder around 2141 cm$^{-1}$, resembling absorption bands previously reported for CO in mixed ice mantles. The high-wavenumber component has been attributed to CO embedded in nonhydrogen-bonding environments \citep{boogert2002}. During deposition, the relative intensity of the 2139 cm$^{-1}$ component progressively increases, reflecting a growing dominance of CO--CO interactions as the pore space becomes filled. The resulting band profile gradually approaches that of multilayer CO.

During warm-up, the two sharp CO peaks on CaF$_2$ (Fig.~\ref{fig3}a) remained spectrally stable up to 29 K, above which both peaks rapidly lost intensity and became undetectable within the experimental sensitivity. This abrupt loss of spectral signal confirms rapid desorption of the CO film, consistent with previous measurements showing multilayer CO desorption near 30 K \citep{van2006}.

On carbonaceous dust, however, CO (Fig.~\ref{fig3}b) exhibited markedly different thermal evolution. Between 18 and 31 K, prior to desorption, the 2139 cm$^{-1}$ component decreased while the 2141 cm$^{-1}$ component increased. These band-profile changes point to thermally activated diffusion into the porous dust matrix. Mobile CO molecules redistribute into deeper adsorption sites associated with higher adsorption energies. Following this diffusion phase, the absorption band remains detectable up to 39 K and declines gradually until 61 K, demonstrating that a substantial fraction of molecules becomes confined within energetically favorable adsorption sites.

\subsection{Carbon dioxide}

In the CO$_2$ $\nu_3$ region, spectra measured on CaF$_2$ (Fig.~\ref{fig2}c) displayed well-resolved LO and TO peaks at 2380 and 2342 cm$^{-1}$, together with a distinct X-band at 2328 cm$^{-1}$. Experimental and theoretical studies showed that LO--TO splitting occurs in amorphous and crystalline CO$_2$ films, whereas the X-band is characteristic of amorphous CO$_2$ and reflects disordered molecular environments \citep{escribano2013}. As deposition proceeded, the X-band weakened while the LO and TO peaks strengthened, pointing to progressive crystallization with increasing film thickness. The nearly constant peak positions throughout the growth reflect a weak coupling between the CO$_2$ ice and the CaF$_2$ substrate.

On carbonaceous dust (Fig.~\ref{fig2}d), CO$_2$ instead formed a single broad absorption band with a weak shoulder that shifted progressively from 2343 to 2349 cm$^{-1}$ during deposition. The absence of clear LO--TO splitting is consistent with spectra reported for isotopically mixed CO$_2$ nanoparticles \citep{bauerecker2005} and early film-growth stages prior to formation of a continuous layer \citep{kumi2006}.

Upon heating, CO$_2$ ice on CaF$_2$ (Fig.~\ref{fig3}c) underwent a thermally driven phase transition. Between 19 and 23 K, the X-band decreased while the LO and TO peaks sharpened and intensified, indicating increasing crystallinity. This band evolution preceded desorption near 78 K, consistent with previous multilayer CO$_2$ measurements \citep{escribano2013}.

By contrast, CO$_2$ deposited on carbonaceous dust (Fig.~\ref{fig3}d) exhibited substantially weaker spectral evolution over the same temperature range, reflecting suppressed crystallization and stronger molecule–surface coupling. The slight band sharpening between 20 and 27 K suggests limited structural relaxation, but this effect remains minor compared with the distinct phase transition observed on CaF$_2$. Between 52 and 68 K, the absorption band shifted from 2346 to 2341 cm$^{-1}$ and became more symmetric, consistent with profiles reported for CO--CO$_2$ mixed solids \citep{van2006}. This evolution points to thermally activated diffusion into the porous dust matrix. As molecular mobility increases, CO$_2$ redistributes into deeper and energetically favorable adsorption sites within the substrate. The CO$_2$ absorption band persisted up to 83 K, and desorption continued until 108 K, substantially higher than on CaF$_2$.

\subsection{Water}

In contrast to CO and CO$_2$, H$_2$O ice exhibited a comparatively weak substrate dependence. On CaF$_2$ (Fig.~\ref{fig2}e), the stretching ($\nu_1$ and $\nu_3$) region exhibited a broad absorption band centered at 3271 cm$^{-1}$ \citep{palumbo2005}, characteristic of amorphous H$_2$O ice \citep{hagen1981}. The absence of substantial band shifts or band profile evolution during deposition reflects weak molecule–surface coupling and stable growth of the hydrogen-bonded network.

The spectra obtained on carbonaceous dust (Fig.~\ref{fig2}f) showed a similarly broad main band centered near 3301 cm$^{-1}$ together with a distinct dangling-OH band at 3702 cm$^{-1}$ \citep{palumbo2005}. The relatively hydrophilic CaF$_2$ supports formation of a continuous hydrogen-bonded network, whereas the hydrophobic carbonaceous surface likely disrupts the classical tetrahedral hydrogen bond network, generating under-coordinated OH groups \citep{rowland1991}. Nevertheless, the overall spectral similarity between the two substrates suggests that intermolecular H$_2$O interactions dominate interactions with the substrate \citep{marchione2019}.

Upon heating, H$_2$O ice on CaF$_2$ (Fig.~\ref{fig3}e) undergoes well-known phase transitions. Between 25 and 80 K, the stretching band shifted from 3271 to 3244 cm$^{-1}$ and narrows, reflecting a transition from loosely bound to more compact amorphous ice \citep{mayer1986, jenniskens1994}. Between 137 and 148 K, the lower-wavenumber side of the band decreased preferentially without substantial peak shifts, consistent with crystallization toward cubic ice \citep{hagen1981}. Above 148 K, the infrared signal rapidly disappeared, signaling desorption of the H$_2$O ice \citep{fraser2001}.

By comparison, H$_2$O deposited on carbonaceous dust (Fig.~\ref{fig3}f) followed a similar thermal pathway. Between 25 and 80 K, the stretching band shifted from 3307 to 3270 cm$^{-1}$ and narrowed slightly, while the dangling-OH band gradually decreased. This trend continued up to 130 K and is consistent with pore collapse and structural compaction during annealing. Between 130 and 143 K, the main band shifted further to 3238 cm$^{-1}$, approaching the position observed on CaF$_2$. A weak residual feature near 3613 cm$^{-1}$ persisted up to 167 K and might correspond to a small population of molecules bound at specific adsorption sites. Unlike CO and CO$_2$, the absence of substantial band broadening or delayed desorption shows that H$_2$O does not significantly diffuse into or become trapped within the porous dust matrix.

The broadening of molecular bands on carbonaceous surfaces can be attributed to the structural and morphological properties of the carbon grains. Their porous structure and large surface area create heterogeneous adsorption environments that suppress cooperative molecular ordering. In addition, the grains contain aromatic and aliphatic subunits. The aromatic component mainly consists of bent graphene layers, with some planar layers. The slight electrical conductivity of the material might also contribute to band broadening through rapid vibrational relaxation via electron–hole pair formation.

\begin{figure}
\resizebox{\hsize}{!}{\includegraphics{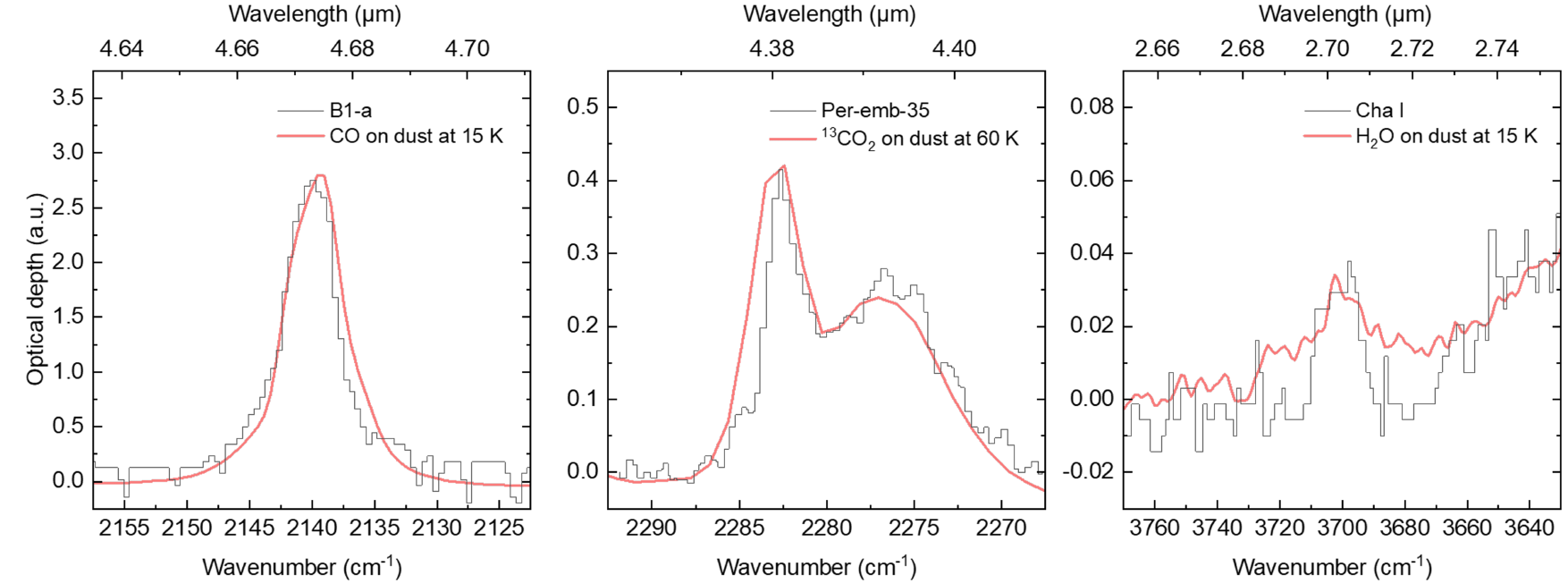}}
\caption{Observed spectra compared to laboratory data for CO (left), $^{13}$CO$_2$ (middle), and H$_2$O (right) ices on carbonaceous dust analogs.}
   \label{fig4}
\end{figure}

\section{Astrophysical implications}

We evaluated the astronomical implications by comparing JWST ice observations with laboratory spectra of CO, CO$_2$, and H$_2$O ices formed on carbonaceous dust analogs. The 4.67 $\mu$m CO stretching band observed toward sources such as \object{B1-a} exhibits substantial variations in the band profile and intensity (Fig.~\ref{fig4}). To facilitate the comparison, we scaled the laboratory spectrum to match the observed intensity. These differences are commonly interpreted as signatures of thermal processing, with the distribution of CO ice primarily governed by its volatility and the temperature gradients along the line of sight \citep{brunken2024}. Our laboratory spectra of CO deposited on porous carbonaceous dust reproduce comparable broadened bands under low-temperature deposition, suggesting that part of the spectral diversity in CO ice might reflect molecule–surface interactions and pore filling within dust mantles.

Observations of CO$_2$ toward the low-mass protostar \object{Per-emb-35} showed complex absorption bands across multiple vibrational modes, including the 4.38 $\mu$m $^{13}$CO$_2$ stretching feature (Fig.~\ref{fig4}). Recently, \citet{suhasaria2025} showed that CO$_2$ deposited on a porous amorphous silicate grain analog developed a split $^{13}$CO$_2$ stretching band above $\sim$60 K, coincident with the onset of molecular diffusion. The resulting spectra closely resemble JWST observations of embedded protostars and provide an alternative to the multicomponent spectral decomposition of astronomical spectra using laboratory components representing distinct molecular environments \citep{brunken2025}. We observed the same behavior on porous amorphous carbon, suggesting that this spectral evolution is a general property of porous interstellar dust analogs. Grain substrates should therefore be accounted for when acquiring and interpreting laboratory spectra for comparison with JWST observations, rather than relying solely on thick substrate-independent ice films.

A dangling-OH feature near 2.703 $\mu$m was observed toward dense clouds such as \object{Cha I} (Fig.~\ref{fig4}). This pattern has been attributed to doubly or triply H-bonded water molecules in the bulk and is often attributed to the porosity of water ice \citep{noble2024}. Our experiments showed that H$_2$O deposited on porous carbonaceous dust grains exhibits persistent substrate-dependent dangling-bond signatures. This dangling-bond feature starts to appear at the dust-surface interface, grows with the thickness of the water ice layer, and eventually extends into the bulk. Variations in dangling-bond signatures might therefore reflect the surface chemistry and morphology of the underlying dust grains, in addition to the bulk ice properties.

Our carbonaceous dust analogs reproduce several key characteristics of observed interstellar ice spectra. Interstellar ice spectra likely encode not only temperature and composition, but also the adsorption landscape and physical structure of the underlying dust.

\section{Conclusions}

We demonstrated that carbonaceous dust exerts a molecule-specific effect on the structure and thermal evolution of interstellar ices. CO and CO$_2$ both readily interact with and diffuse into the porous dust matrix, producing broadened and redshifted infrared absorption bands together with delayed thermal desorption. In contrast, H$_2$O remains largely decoupled from the substrate and retains its spectral and thermal properties across both surfaces.

Our results establish the necessity of incorporating molecule–surface interactions into astrochemical models, as neglecting such effects may lead to misinterpretations of infrared observations, particularly in dust-rich environments such as dense cores and protoplanetary disks, where ice bands often deviate from the profiles expected for thick substrate-independent films. The observed desorption dynamics further constrain models of gas–solid exchange and volatile redistribution in star-forming regions.

This information is essential for interpreting ice spectra from current and upcoming observatories, including the JWST, and for maximizing the scientific return of missions. Future work should extend this framework to additional astrophysically relevant species and examine how substrate composition and nanostructure control adsorption, reactivity, and desorption.

\begin{acknowledgements}

This work was supported by National Science and Technology Council (NSTC), Taiwan under grant nos. NSTC 110-2923-M-008-004-MY3, NSTC 114-2112-M-008-017, and NSTC 114-2639-M-A49-002-ASP (Y.-J.C.). T.S. and C.J. are grateful to Deutsche Forschungsgemeinschaft (DFG) for support under grant JA 2107/11-1 (project No. 504825294) and JA 2107/10-1 (project No. 468269691), respectively.

\end{acknowledgements}

\bibliographystyle{aa}
\bibliography{Reference}
\end{document}